# Combined Image Encryption and Steganography Algorithm in the Spatial Domain

Aya H. S. Abdelgader[1], Raneem A. Aboughalia[2], Osama A. S. Alkishriwo[3]

[1] A.Abdelgader@uot.edu.ly, [2] Raneem.abg@gmail.com, [3] alkishriewo@yahoo.com

[1, 2, 3] Department of Electrical and Electronic Eng., College of Eng., University of Tripoli, Libya

*Corresponding author email: alkishriewo@yahoo.com

**ABSTRACT**

In recent years, steganography has emerged as one of the main research areas in information security. Least significant bit (LSB) steganography is one of the fundamental and conventional spatial domain methods, which is capable of hiding larger secret information in a cover image without noticeable visual distortions. In this paper, a combined algorithm based on LSB steganography and chaotic encryption is proposed. Experimental results show the feasibility of the proposed method. In comparison with existing steganographic spatial domain based algorithms, the suggested algorithm is shown to have some advantages over existing ones, namely, larger key space and a higher level of security against some existing attacks.

## 1   Introduction

The growing of digital communication technologies has caused a substantial increment in data transmission. When sensitive information such as bank account numbers is being shared between two communicating parties over a public channel, security of such data becomes necessary. Cryptography and steganography are two important tools for providing security and protecting sensitive information. Cryptography provides features such as confidentiality and integrity of data. For instance, confidentiality is achieved via an encryption algorithm which scrambles/mixes the private information so that it becomes unreadable to any party other than the intended recipient. However, steganography provides data security by hiding the information in a cover medium so that even the existence of a hidden message is not known to an intruder. Secret messages are embedded in cover objects to form stego objects. These stego objects are transmitted through the insecure channel. Cover objects may take the form of any digital image, audio, video and other computer files. Digital images are widely used as cover object of hidden information because of the high level of redundancy in them which is caused by the low sensitivity of the human visual system to details. The success of steganography lies in transmission of stego objects without suspicion [1].

A large number of image steganographic techniques have appeared in the literature, for example [2-7]. These techniques can be classified into two main classes: spatial domain and transform domain techniques. In spatial domain techniques, private message is embedded in the intensity of image pixels directly [2-4]. In transform domain techniques, the private message is embedded in the cover by modifying coefficients in a transform domain such as discrete cosine transform (DCT) and integer discrete wavelet transform [5-6].

Although transform domain based algorithms are more robust to steganalytic attacks, the spatial domain based algorithms such as least significant bit (LSB) algorithms are much simpler and faster. Several versions of the LSBs embedding algorithms have appeared in the literature. However, many steganalysis tools that reveal the insecurity of some LSBs replacement algorithms have been reported. For example, in [7] authors suggested a steganalytic attack that can estimate the length of information embedded in a host image for various LSBs algorithms. Nevertheless, the high embedding capacity and low computational complexity of these algorithms have encouraged researchers to further participate in this area.

Chaotic maps are well known for their sensitivity to initial conditions and control parameters. These properties make them suitable for building blocks in the design of many cryptographic and steganographic algorithms [3, 8]. In this paper, we propose a new LSBs spatial domain algorithm that is based on mixing two 2-D chaotic maps. The proposed algorithm encrypts the secret message using mixed chaotic map and uses LSB for data hiding.

The rest of the paper is organized as follow: Section 2 presents the used 2-D chaotic maps. In Section 3, we give a detailed description of the proposed algorithm and a flowchart. In Section 4, simulation results are presented and discussed. The conclusions are given in Section 5.

## 2  Two Dimensional Chaotic Maps

In the proposed steganography method, we have used a combination of two 2D chaotic systems which are logistic and duffing maps defined in [8, 9] as given in (1) and (2).

$$\begin{aligned} x_{n+1} &= \mu\, x_n\,(1 - x_n) \\ y_{n+1} &= \lambda\, y_n\,(1 - y_n) \end{aligned} \quad (1)$$

where, $\mu$, $\lambda$, $x$ and $y$ are the control parameters and state values, respectively. When $\mu$ and $\lambda \in [3.57, 4]$, the system is chaotic. The Duffing map depends on the two constants $a$ and $b$. These are usually set to $a = 2.75$ and $b = 0.2$ to produce chaotic behaviour. It is a discrete version of the Duffing equation.

$$\begin{aligned} z_{n+1} &= w_n \\ w_{n+1} &= -b\, z_n + a w_n - w_{n+1}^3 \end{aligned} \quad (2)$$

## 3   The Proposed Steganographic Algorithm

The steganographic scheme proposed in this article embeds a binary message according to the least significant bit technique as shown in Figure 1. This helps imperceptibility since the more significant bits of the cover image are not altered. Data embedded is done using the following steps:

- Step 2: Read both of cover image and secret image, the cover image most be equal or larger than the secret image.
- Step 3: use chaotic maps to encrypt secret image.
- Step 3: Select the block size for the encryption algorithm and generate random initial conditions for the chaotic maps.
- Step 4: Using the initial conditions to generate chaotic maps key streams $X$ and $Y$.
- Step 5: Secret image is divided into blocks of same size ($m \times m$), scrambled using the encryption key stream and recombined into a single image.
- Step 6: Pixel wise XOR operation is done on the scrambled image using the key stream to get the encrypted image.
- Step 7: Extract the pixels of the cover image and encrypted secret image.
- Step 8: Choose first pixel of the cover image and pick first pixel of the encrypted secret image then place it using LSB algorithm, one pixel of the encrypted secret image have 8 bits, using for example 8bpp all this bits will be hidden inside one pixel of the color image.
- Step 10: Repeat step 9 till all the pixels of the encrypted secret image has been embedded.

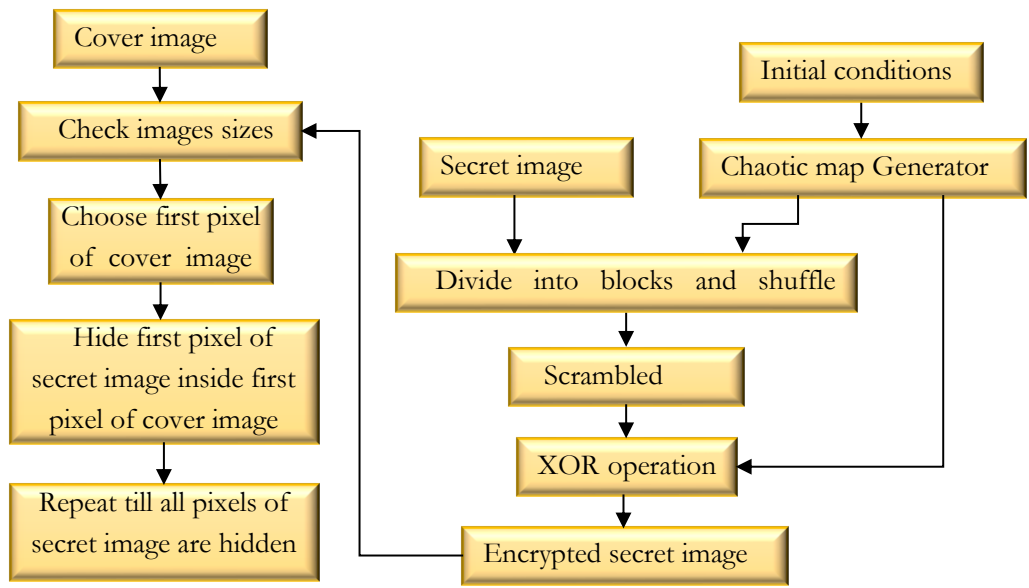

**Figure 1:** *Block diagram of proposed steganography algorithm.*

When applying LSB techniques to each byte of a 24 bit image, we can take the binary representation of the hidden data and overwrite the LSB of each byte within the cover image. If the LSB of the pixel value of cover image $C(i,j)$ is equal to the next message bit $SM$ of secret massage to be embedded, $C(i,j)$ remain unchanged; if not, set the LSB of $C(i,j)$ to $SM$.

## 4 Performance Analysis and Experimental Results

In this section, experimental results are given to demonstrate the performance of the proposed algorithm. Comparative experimental studies are also presented to show the superiority of the proposed algorithm over typical existing ones. Four standard $512 \times 512 \times 3$ colored images, namely, Airplane, Fruits, pool, and girl are used as cover images for hiding sensitive information of length $524288$ bit.

### 4.1 Visual Attack

Visual attacks, regarded as the simplest type of steganalysis, aim at revealing the presence of hidden information through visual inspection by the naked eye. The presented algorithm is designed to be robust against visual attacks. Figure 2 presents a cover image ($512 \times 512 \times 3$ Airplane), a secret-image carrying of size ($256 \times 256$), an encrypted secret image, and a stego-image carrying an encrypted secret image. A visual inspection of the cover and the stego-image does not reveal any difference between the two images.

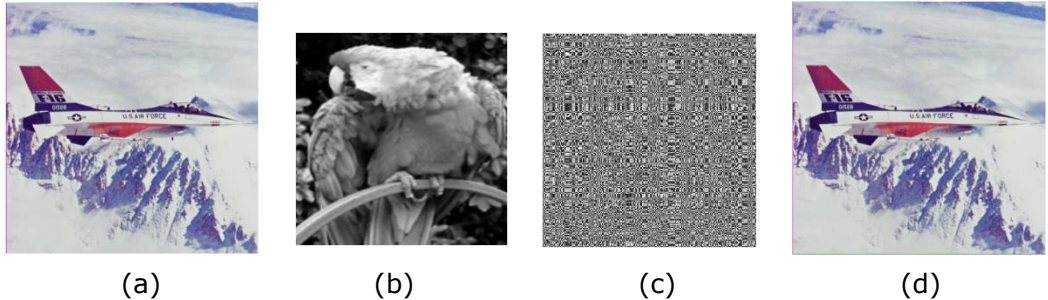

(a)   (b)   (c)   (d)

**Figure 2:** *(a) cover image, (b) secret image, (c) encrypted secret image, (d) stego image.*

### 4.2 Imperceptibility and Payload

For data hiding in images, hiding capacity and visual quality of the scheme play important roles. So, increasing hiding capacity adversely affects the visual quality of the stego-image. The embedding rate is the number of bits that can be embedded into one pixel, and it is measured by bits per pixel (bpp). It is known that human visual system cannot detect the distortion of a stego-image, when the peak signal to noise ratio (PSNR) is higher than $30$ dB.

To compare between each of 3, 6, 8 bits per pixel, we measure PSNR for all stego-images as shown in Table 1, the highest PSNR values means the stego-image is similar to cover image.

**Table 1:** *PSNR comparison in dB.*

| bpp | Airplane | Fruits | Pool | Girl |
|---|---|---|---|---|
| 3 | 53.0 | 53.6 | 53.8 | 53.6 |
| 6 | 48.2 | 48.5 | 48.2 | 48.6 |
| 8 | 41.5 | 41.8 | 42.7 | 41.5 |

In Table 2, PSNR (dB) is calculated with different payload capacity of 3 bpp on a stego-image using Lena as a cover image, and the results are compared with similar steganography schemes for the same cover image.

**Table 2:** Comparison of the proposed algorithm to existing work in terms of PSNR (dB).

| bpp | Proposed | [4] | [10] | [11] |
|---|---|---|---|---|
| 3 | 53.0 | 37.9 | 37.3 | 37.8 |

### 4.3 Image Histogram

In Figure 3, we present the histograms of the cover image Lena and the resulting stego-image produced by our algorithm with a message of size 3 bpp, 6 bpp, and 8 bpp. It can be observed that the two histograms are very similar. This test shows a comparison between the cover image and the stego image, using the histogram as a visual comparison tool.

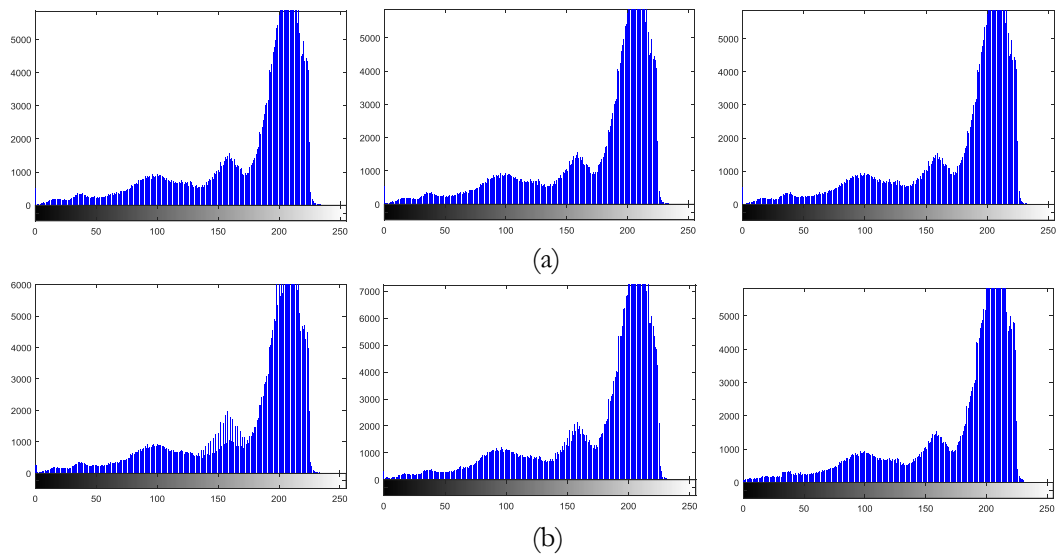

**Figure 3:** (a) *Histogram of cover image, (b) Histogram of stego-image from right to left 3 bpp, 6 bpp, and 8 bpp.*

**4.4 Key Space Analysis**

The key space of an encryption algorithm should be large enough to resist brute–force attacks. In the proposed algorithm, the secret key contains seven real numbers (two control parameters and four initial states). If we assume the computation precision of the computer is $10^{-15}$, then the key space is about $10^{90} \approx 2^{299}$. Such a large key space can ensure a high security against brute–force attacks.

**5 Conclusions**

In this paper, a steganographic algorithm based on two 2-D chaotic maps has been introduced. This algorithm embeds the encrypted sensitive information using chaotic maps into the cover image according to the least significant bit technique. The LSB algorithm effectively allows the embedding of secret information at higher level frequencies, which are not visible to the human eye. The presented simulation results show the resistance of the suggested algorithm against some existing steganalytic attacks. Furthermore, the results show its advantages over some existing algorithms.